\documentclass[epj]{svjour}
\usepackage{graphicx,bm,amsmath,amssymb}
\usepackage[a-1b]{pdfx}
\hypersetup{colorlinks=true,
  urlcolor=blue,       
  filecolor=green,     
  linkcolor=red}
  
\def\today{\number\day\space\ifcase\month\or
January\or February\or March\or April\or May\or June\or
July\or August\or September\or October\or November\or December\fi
\space\number\year}
\newcount\mins \newcount\hours
\def\now{\hours=\time \mins=\time
	\divide\hours by60 \multiply\hours by60 \advance\mins by-\hours
	\divide\hours by60 
	\number\hours:\ifnum\mins<10 0\fi\number\mins~}

\usepackage{xcolor}

\newcommand{\subrm}[1]{{\scriptscriptstyle\mathrm{#1}}}
\newcommand{\bfnab}{\bm{\nabla}}
\newcommand{\bfnabl}{\raisebox{0.1em}{$\,\stackrel{\scriptstyle\leftarrow}\bfnab$}}

\begin{document}
\title{Quark flavor physics and lattice QCD}
\author{Matthew Wingate
}                     
%
%
\institute{DAMTP, University of Cambridge, Wilberforce Road, Cambridge, CB3 0WA,  UK\thanks{\email{\href{mailto:M.Wingate@damtp.cam.ac.uk}{M.Wingate@damtp.cam.ac.uk}}}}
\date{22 June 2021}
%
\abstract{
For a long time, investigation into the weak interactions of quarks has guided us toward understanding the Standard Model we know today.  Now in the era of high precision, these studies are still one of the most promising avenues for peering beyond the Standard Model.  This is a large-scale endeavour with many tales and many protagonists.  In these pages I follow a few threads of a complex story, those passing through the realm of lattice gauge theory. 
%
%
} 
\maketitle
%

\section{Introduction}
\label{sec:intro}

In the Standard Model (SM),  flavor-changing interactions are mediated by $W$ bosons.  Electroweak symmetry breaking gives mass to the quarks and, in doing so, induces mixing between the $\mathrm{SU}(2)_L$ doublets.  The relation between weak eigenstates and mass eigenstates is given by the Cabibbo-Kobyashi-Maskawa (CKM) matrix
\begin{align}
V_{\mathrm{CKM}} = \begin{pmatrix}
V_{ud} & V_{us} & V_{ub} \\
V_{cd} & V_{cs} & V_{cb} \\
V_{td} & V_{ts} & V_{tb} 
\end{pmatrix} \,.
\end{align}
Taking into account the requirement of the unitarity of $V_{\mathrm{CKM}}$ and the phase-invariance of the quark fields, there are four independent parameters governing quark flavor-changing interactions.

There could be more to the story, however.  Is there a reason why electroweak symmetry breaking produces a light scalar boson with a mass just so?  Is there another undiscovered source of CP violation in the quark sector which could explain why matter dominates antimatter in the universe? Is there a particle which could make up the dark matter inferred from astrophysical observations?  Many of the ``Beyond the Standard Model (BSM)'' models addressing these questions could affect quark flavor interactions.

By making a plethora of measurements with increasing precision, particle physicists hope to  constrain the four independent CKM parameters so tightly that an inconsistency emerges, a gap that could only be explained by BSM physics.  Because experiments measure the weak interactions of hadrons, the bound states of quarks, precise QCD calculations are required to draw inferences about quark interactions from these measurements.  This is where lattice QCD plays an important role, one which I aim to review here.

In the pages that follow, I will focus on a few stories rather than attempt an encyclopedic account.  In studying these stories, I was struck by the emergence of some common features, which in turn reminded me of the notion of a ``monomyth'' or ``Hero's Journey,'' popularized by Joseph Campbell in the late 1980's.  Literary work fitting this template includes \textit{The Iliad} and \textit{The Odyssey}, \textit{Moby Dick}, and \textit{Jane Eyre}. At the time Campbell illustrated the theory with \textit{Star Wars} as his main example.   Tolkien's \textit{The Hobbit} and \textit{Lord of the Rings} also follow the same arc.

\begin{figure}
\centering
\resizebox{0.75\hsize}{!}{\includegraphics{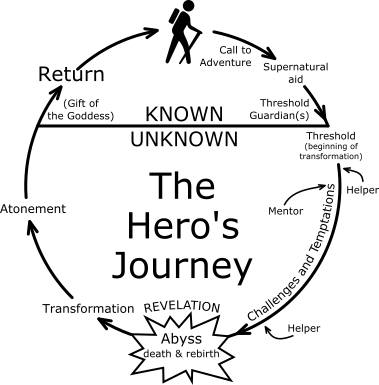}}
\caption{\label{fig:hero}The sagas of quark flavor heroes are classic tales of adventure. [Image source: 
\href{https://commons.wikimedia.org/wiki/File:Heroesjourney.svg}{Wikimedia Commons}, public domain.]}
\end{figure}

Figure \ref{fig:hero} depicts some key points in The Hero's Journey.  Many of these resonate with the adventures of flavor physics heroes.
It all starts with an idea for a new measurement, a new calculation, a new BSM signature.  A fellowship must be formed.  A proposal must be written in order to satisfy the gatekeepers; this often requires good luck or other supernatural aid.  The flavor hero cannot get far into the unknown without the aid of helpers and mentors, be they technicians, accelerator experts, research software engineers, or otherwise.  Many challenges must be faced -- bugs, downtime, statistical and systematic errors -- and temptations must be resisted -- premature publication, under- (or over-)estimating uncertainties.  Finally the innermost cave is reached, where scientific progress is made, error bars reduced, tensions made or released.  The flavor hero is reborn and must carry their revelation back to the known world.  Results must be interpreted and explained.  A sacrificial act of atonement must be made to satisfy peer reviewers. Finally the hero can rest and contemplate new adventures.

In these pages I tell a few tales of these adventurers.  In the tradition of oral storytelling, this version will be different from the ones told by others, at least in the details. Nevertheless out of the many tellers telling tales, a saga emerges.

\section{CKM}
\label{sec:ckm}

Most of the flavor physics stories are set in the square of the CKM matrix, with its broad avenue down the diagonal and its two darker corners on either side.  As long as quark flavor measurements consistently agree with the CKM parametrization, the Standard Map of weak interactions is sufficient to capture everything.  After decades of refining the map, it is quite detailed now.  Flavor heroes have to work very hard to try to find discrepancies.

We can organize the various adventures according to paths through the map, three east-to-west and one north-to-south.  I will include several details, but a much more detailed review is available from the Flavour Lattice Averaging Group (FLAG) \cite{Aoki:2019cca} (also see their 2020 web update \cite{FLAG:2020update}).

In order to be accurate beyond 10-20\%, lattice QCD calculations must include the effects of light sea quarks \cite{Davies:2003ik}.  Most modern work also includes a dynamical strange quark, with its mass tuned close to the physical value. The calculations I discuss below have either $2+1$ or $2+1+1$ flavors of sea quarks, the difference being whether charm sea quark effects are included.

For quark masses at scales where $\alpha_s(m_q)$ is small, one can estimate the effects of heavy sea quarks perturbatively.  Expanding the fermion determinant in inverse powers of $m_q$, one finds that contributions from charm quark loops are of the order $\alpha_s(\Lambda_{\subrm{QCD}}/2m_q)^2$ \cite{Nobes:2005yh}.  For the charm quark, this is at the 1-2\% level (e.g.\ \cite{Bazavov:2016nty}).  A recent study looking at the charmonium spectrum in theories with either 0 or 2 dynamical charm-like quarks (and no other sea quarks) found quenching effects in agreement with the perturbative estimate \cite{Cali:2019enm}.   Therefore, lattice results with 2+1 flavors of sea quarks, i.e.\ those which omit the effect of sea $c$ quark loops, can still provide important information in cases where other errors are dominant.

\subsection{First row unitarity}
\label{ssec:firstrow}

Unitarity of the CKM matrix implies that $|V_{ud}|^2 + |V_{us}|^2 + |V_{ub}|^2 = 1$.
This section is really about $|V_{ud}|$ and $|V_{us}|$ since $|V_{ub}|$ is so small.  The important tale of $|V_{ub}|$ will be told later.

The matrix element $|V_{ud}|$ is most precisely determined through superallowed nuclear $\beta$ decays \cite{Hardy:2014qxa}.  Recent reevaluations of radiative corrections \cite{Seng:2018yzq,Czarnecki:2019mwq,Seng:2020wjq} have shifted the central value for $|V_{ud}|$ down by $2\sigma$ to $|V_{ud}| = 0.97370(14)$ compared to the 2018 PDG value \cite{Tanabashi:2018oca,Zyla:2020zbs}.  Consequences of this shift are still being studied \cite{Seng:2018qru,Gorchtein:2018fxl}, so there may be more to the story.  A test of first-row unitarity crucially depends on $|V_{ud}|$, given its relative size.

$|V_{ud}|$ can also be inferred from neutron $\beta$ decay.  This relies on precise knowledge of the neutron lifetime, the ratio of axial vector to vector couplings, $g_A$, and the same electroweak radiative corrections discussed above.  There is presently some disagreement in experimental measurements of the neutron lifetime depending on whether it is determined in beam experiments or with trapped ultracold neutrons \cite{Pattie:2017vsj}.  In principle lattice QCD could contribute with a determination of $g_A$, but the experimental measurement \cite{Markisch:2018ndu} is more precise by a factor of about 50 than lattice results \cite{Chang:2018uxx,Liang:2018pis,Gupta:2018qil}.  In Ref.~\cite{Czarnecki:2018okw} an argument is made to prefer the lifetimes from trapped neutron experiments. Taking their average for the mean lifetime, the recent result for $g_A$  \cite{Markisch:2018ndu}, and the new radiative correction  \cite{Seng:2018yzq} leads to a $|V_{ud}|$ from neutron decay of 0.97377(78), an uncertainty 5 times larger than from the nuclear decays.

Pion $\beta$ decay, $\pi^+ \to \pi^0 e^+ \nu_e$, can also tell us about $|V_{ud}|$.  The hadronic form factor at zero recoil is equal to 1 within the accuracy needed here, so lattice QCD is not needed to provide a normalization.  Lattice calculations can help with the radiative correction factors though \cite{Feng:2020zdc}. 

We cannot fully answer the unitarity question while there is a new knot to untangle in the $|V_{ud}|$ story.  Nevertheless there is another unresolved plot-line in the first row, one where kaons are the main characters.

\subsubsection{Decay constants}

\begin{figure}
\centering
\resizebox{0.95\hsize}{!}{\includegraphics{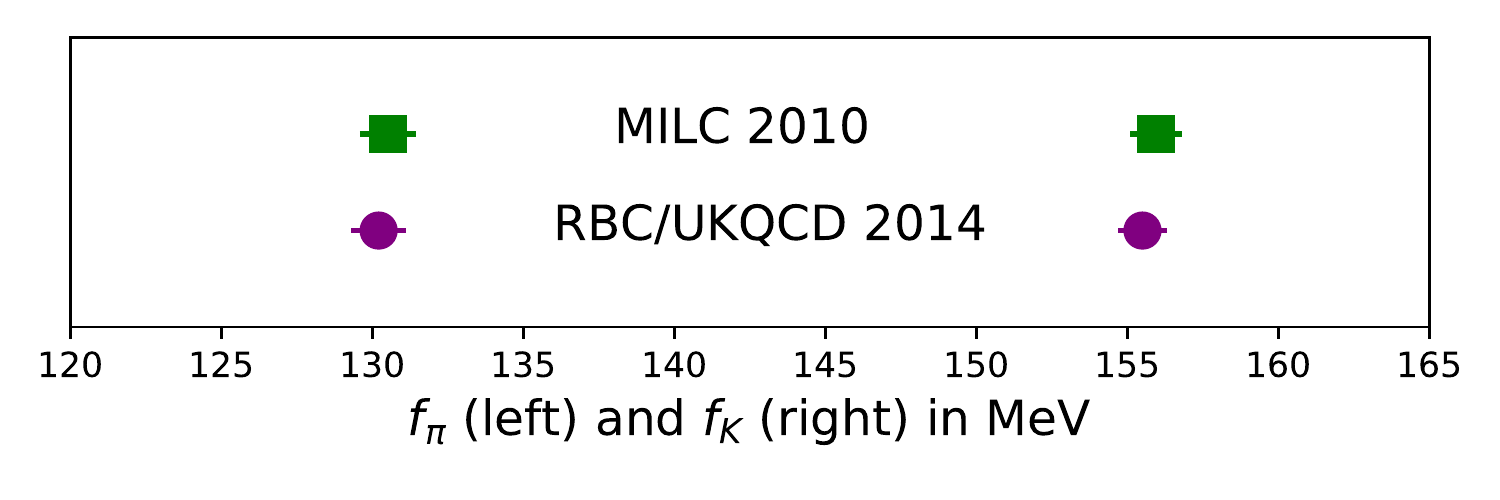}}
\caption{\label{fig:fK_fpi_nf21_compare}Comparison of decay constants at the sub-percent level, on two independent $n_f = 2+1$ ensembles using completely different lattice discretizations \cite{Bazavov:2010hj,Blum:2014tka}.}
\end{figure}

The ratio $|V_{us}/V_{ud}|$ can be precisely determined from measurements of the leptonic decays $K\to \mu \nu$ and $\pi \to \mu \nu$, combined with lattice QCD determinations of the decay constants $f_K$ and $f_\pi$ \cite{Marciano:2004uf}.
FLAG \cite{Aoki:2019cca,FLAG:2020update} provides an extensive summary of many results for the decay constants. There is good agreement among results, even comparing $2+1$ flavor \cite{Follana:2007uv,Durr:2010hr,Bazavov:2010hj,Blum:2014tka,Durr:2016ulb,Bornyakov:2016dzn}
to $2+1+1$ \cite{Dowdall:2013tga,Carrasco:2014poa,Bazavov:2017lyh,Miller:2020xhy}.
After accounting for strong isospin the FLAG averages are
\begin{align}
\label{eq:fK_fPi_FLAG}
 f_{K^\pm}/f_{\pi^\pm} & = 1.1932(21) &  n_f & = 2+1+1 \\
f_{K^\pm}/f_{\pi^\pm} & = 1.1917(37) & n_f & = 2+1 \,.
\end{align}
The agreement between individual lattice results is noteworthy considering the variety of lattice discretizations used.  Especially interesting is the comparison of $2+1$ flavor results shown in Fig.~\ref{fig:fK_fpi_nf21_compare}, using the staggered \cite{Bazavov:2010hj} vs.\ domain wall formulations \cite{Blum:2014tka}.   This is a head-to-head test on independent gauge field configurations, using different fermion and gauge discretizations, with both calculations reaching uncertainties as small as $0.5\%$.  They agree perfectly.

Very recently, the ETM Collaboration announced a new result for $f_K/f_\pi$ \cite{Alexandrou:2021bfr}, an update of their previous $2+1+1$ flavor result \cite{Carrasco:2014poa} with the uncertainty reduced from $1.4\%$ to $0.4\%$. Since the other results in the FLAG average share gauge field ensembles with staggered quarks \cite{Dowdall:2013tga,Bazavov:2017lyh,Miller:2020xhy} and staggered valence quark actions \cite{Dowdall:2013tga,Bazavov:2017lyh}, it is very nice to have a precise result with an independent ensemble and a different fermion formulation, twisted-mass fermions in this case.

It appears from (\ref{eq:fK_fPi_FLAG}) that any effects due to quenching the charm quark cancel in the ratio of decay constants, at least at the few per-mille level.  Nevertheless, in discussion below, I will use the more precise 2+1+1 FLAG average.

The precision of the QCD matrix element is now comparable to the expected size of QED and other isospin breaking effects. Radiative corrections have recently been calculated on the lattice \cite{Giusti:2017dwk} and agree with the estimate from chiral perturbation theory \cite{Cirigliano:2011tm}.

\subsubsection{Semileptonic decay}

In order to infer $|V_{us}|$ from semileptonic decays $K\to \pi \ell \nu$, lattice QCD determination of the form factor $f_+(q^2)$ is required.  In fact, the kinematic dependence is fit by each experiment separately and integrated, so all that is needed is the normalization $f_+(0)$.
 Experiments find consistent $|V_{us}|f_+(0)$ for these decays for charged and neutral kaons, and for electron and muon final states, i.e.\ for $K^\pm_{e3}$, $K^\pm_{\mu 3}$, $K^L_{e3}$, $K^L_{\mu 3}$, $K^S_{e3}$ (Table 66.1 of \cite{Zyla:2020zbs}).

\begin{figure}
\centering
\resizebox{0.95\hsize}{!}{\includegraphics{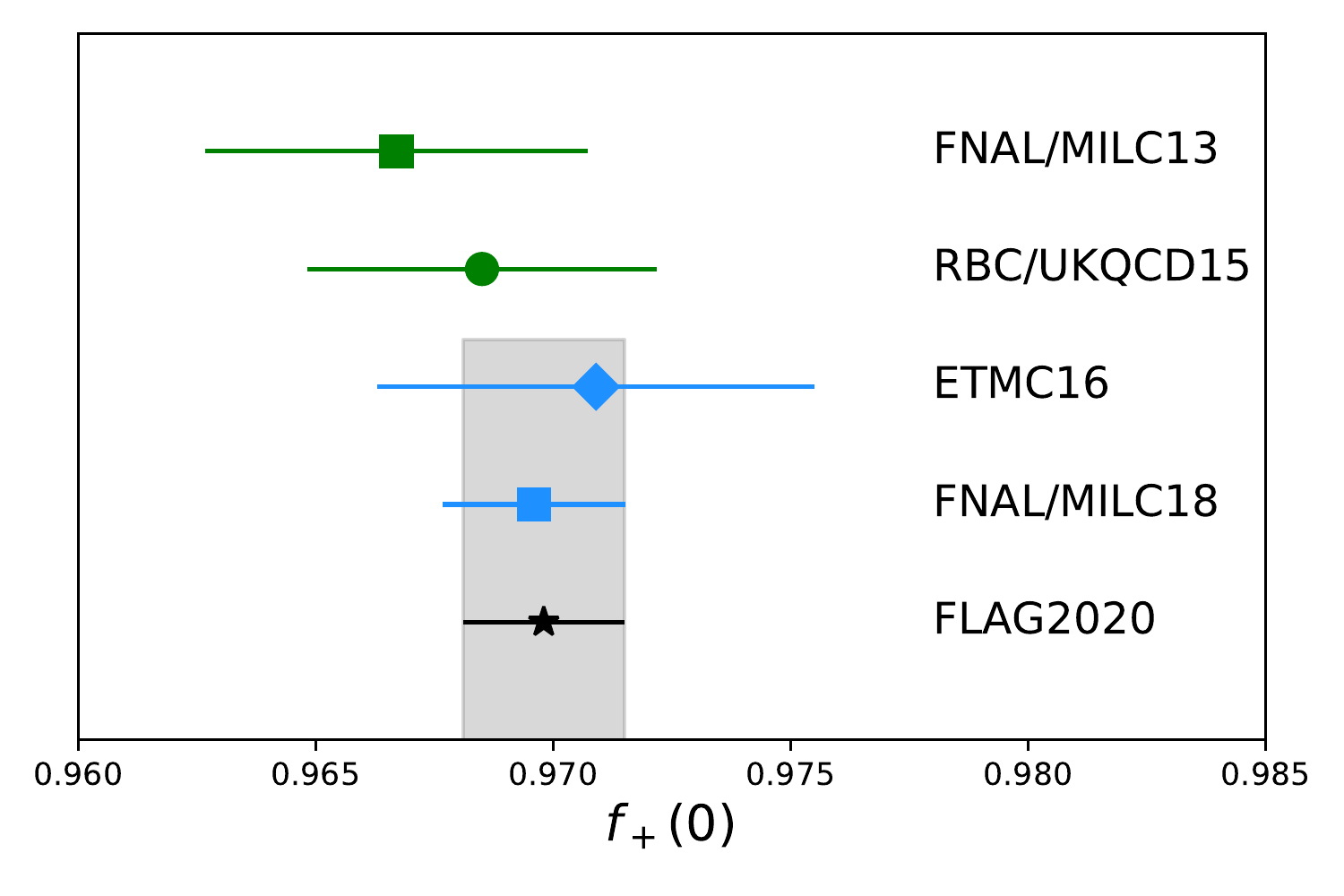}}
\caption{\label{fig:fplus0_compare} Recent results for $f_+(0)$.  The top two results do not include the effects of a dynamical charm quark while the next two do.  The black star and grey band show the FLAG2020 average \cite{FLAG:2020update} of the $n_f=2+1+1$ flavor results. The calculations use staggered (squares) \cite{Bazavov:2012cd,Bazavov:2018kjg}, domain wall (circle) \cite{Boyle:2015hfa}, and twisted mass (diamond) \cite{Carrasco:2016kpy} fermion actions.}
\end{figure}

Lattice results are available with $n_f=2+1$ flavors of sea quarks \cite{Bazavov:2012cd,Boyle:2015hfa} and with 
2+1+1 flavors \cite{Carrasco:2016kpy,Bazavov:2018kjg}. (Ref.\ \cite{Bazavov:2013maa} is superceded by \cite{Bazavov:2018kjg}.)
Any average is dominated by the Fermilab/MILC result \cite{Bazavov:2018kjg}.  FLAG quote  \cite{FLAG:2020update}
\begin{align}
f_+(0) & = 0.9698(17) \,.
\label{eq:fplus0_flag}
\end{align}
 Bearing in mind that the nontrivial part of the calculation is the difference $1 - f_+(0)$, presently determined with a 5\% uncertainty, the effect of quenching the charm quark is not expected to be significant here, as supported by the agreement between $2+1$ and $2+1+1$ flavor results (Fig.~\ref{fig:fplus0_compare}).

There are also some recent results which are on their way to meeting the FLAG criteria for inclusion in their averages. JLQCD has studied the quark mass dependence of $f_+(0)$ using the overlap formulation for the quarks, so far with just a single value of the lattice spacing \cite{Aoki:2017spo}.  Their result is consistent with (\ref{eq:fplus0_flag}).
 PACS has performed a calculation with physical quark masses ($n_f= 2+1$) on a large volume at a single lattice spacing \cite{Kakazu:2019ltq}.  They use an improved Wilson fermion discretization. Their result, after combining their errors is $f_+(0) = 0.960(5)$, where the largest uncertainty is due to discretization errors.   As they continue to finer lattice spacings, it will be interesting to see if their central value remains low compared to the average (\ref{eq:fplus0_flag}).

As will be clear in the next section, there is renewed scrutiny being placed on the SM prediction of $|V_{us}|$ from semileptonic $K$ decay.  One area yet to be addressed are radiative corrections.  Work has begun extending what has been done for $\pi^+ \to \pi^0 e^+ \nu_e$ \cite{Feng:2020zdc} to $K\to \pi \ell \nu$ \cite{Seng:2020jtz}.

\subsubsection{Summary}

\begin{figure}
\centering
\resizebox{0.95\hsize}{!}{\includegraphics{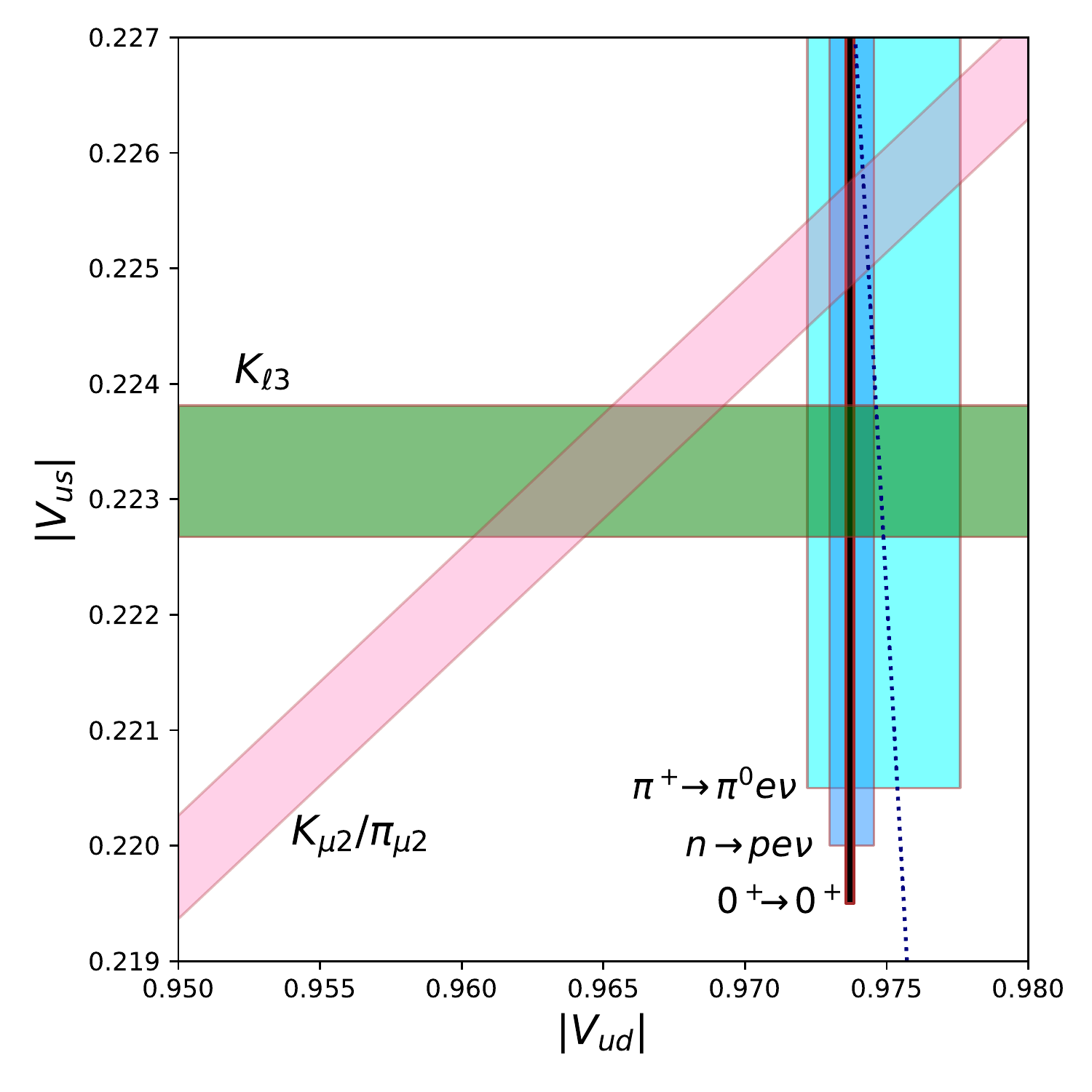}}
\caption{\label{fig:VusVud} Constraints on CKM matrix elements $|V_{ud}|$ and $|V_{us}|$.  The horizontal band is the constraint from semileptonic $K\to \pi \ell \nu$ decay ($K_{\ell 3}$) using (\ref{eq:fplus0_flag}) for $f_+(0)$ and the average of experimental results $|V_{us}| f_+(0) = 0.2165(4)$ \cite{Moulson:2017ive}.  The diagonal band is the constraint from the ratio of leptonic decay rates of the $K$ or $\pi$ to $\mu \nu_\mu$ ($K_{\mu2}/\pi_{\mu2}$), $|V_{us}/V_{ud}| f_{K^\pm}/f_{\pi^\pm} = 0.27600(37)$ \cite{Zyla:2020zbs}, and (\ref{eq:fK_fPi_FLAG}).  The vertical bands are, from narrowest to widest, the constraints from superallowed nuclear decays, neutron $\beta$ decay, and $\pi^+ \to \pi^0 e^+\nu_e$.  The dotted curve indicates values of $|V_{ud}|$ and $|V_{us}|$ consistent with first row unitarity.
}
\end{figure}

The heroes of first-row flavor physics have sent home some tantalizing puzzles.  This goes to show that even well-trod paths such as $\beta$ decays and (semi)leptonic  $\pi$ and $K$ decays can lead to previously uncovered mysteries.  With the scales of uncertainty falling from our eyes, a tension is revealed in the $|V_{ud}|$-$|V_{us}|$ plane (Figure~\ref{fig:VusVud}).

There are really two questions to be asked regarding the constraints in Figure~\ref{fig:VusVud}.  First, are two CKM parameters enough to describe the interactions of $u$ quarks with $d$ and $s$ quarks?  Only if the answer is yes can we ask the second question, is the global fit to $|V_{ud}|$ and $|V_{us}|$ consistent with CKM unitarity?

The answer to the first question is clearly ``no'' at present.  There is no way to obtain a satisfactory fit to the three classes of constraints shown.  What is also clear is a consistency in the $|V_{ud}|$ determinations; even with the disparate ranges of uncertainties, it seems implausible that some change in theory or experiment would resolve the tension by resulting in a $|V_{ud}|$ consistent with the intersection of the $K_{\ell 3}$ and $K_{\mu2}/\pi_{\mu2}$ constraints.  The question is then whether further investigation will move the diagonal band down or the horizontal band up, or neither.

If we assume that the first row is dominated by SM physics, even at the very precise level we have reached, then unitarity constraints hint that the $K_{\ell 3}$ constraint is too low.  Thus there is good cause to look more deeply into the assumptions being made in those determinations.

\subsection{Second row}

In order to test second row CKM unitarity, primary focus lies on $|V_{cd}|$ and $|V_{cs}|$.
$|V_{cb}|$ is too small to be important at the current level of precision.  The quest for $|V_{cb}|$ is the subject of a later section.

\subsubsection{Leptonic decays}

A summary of experimental measurements of $D_{(s)}$ leptonic decays is given in \S71.3 of \cite{Zyla:2020zbs}.  
One notable change from the previous version \cite{Tanabashi:2018oca} is that Sirlin's electroweak correction \cite{Sirlin:1981ie} has now been applied in inferring $|V_{cq}| f_{D_q}$ from the experimental branching fractions.  This significantly reduces tension in the second row unitarity tests, as we will see below.

There are two independent 2+1+1 flavor results, one by ETM \cite{Carrasco:2014poa} and the other by Fermilab/MILC \cite{Bazavov:2017lyh}.  They are in good agreement, though the precision of the latter dominates any average. It is worth mentioning the good agreement seen between different methods on configurations with 2+1 flavors of sea quarks.   Particularly impressive is the agreement at the 1-2\% level between the staggered computations \cite{Davies:2010ip,Na:2012iu} and the completely independent results using domain wall fermions \cite{Boyle:2017jwu}. A  result for $f_{D_s}$ using overlap valence fermions on the $n_f=2+1$ RBC/UKQCD domain wall configurations is also in good agreement \cite{Yang:2014sea}.

An updated result for the ratio $f_{D_s}/f_D$ has been obtained using the domain wall formulation for all quarks \cite{Boyle:2018knm}.  Another new calculation of charmed and $\phi$ meson decay constants \cite{Chen:2020qma}, obtained using overlap valence fermions on the RBC/UKQCD $n_f = 2+1$ domain wall configurations has recently appeared, although the uncertainties are not yet as precise as those above.

\subsubsection{Semileptonic decays}

 Experimental data for semileptonic $D$ decay is summarized by HFLAV \cite{Amhis:2019ckw}.

\begin{figure}
\centering
\resizebox{0.95\hsize}{!}{\includegraphics{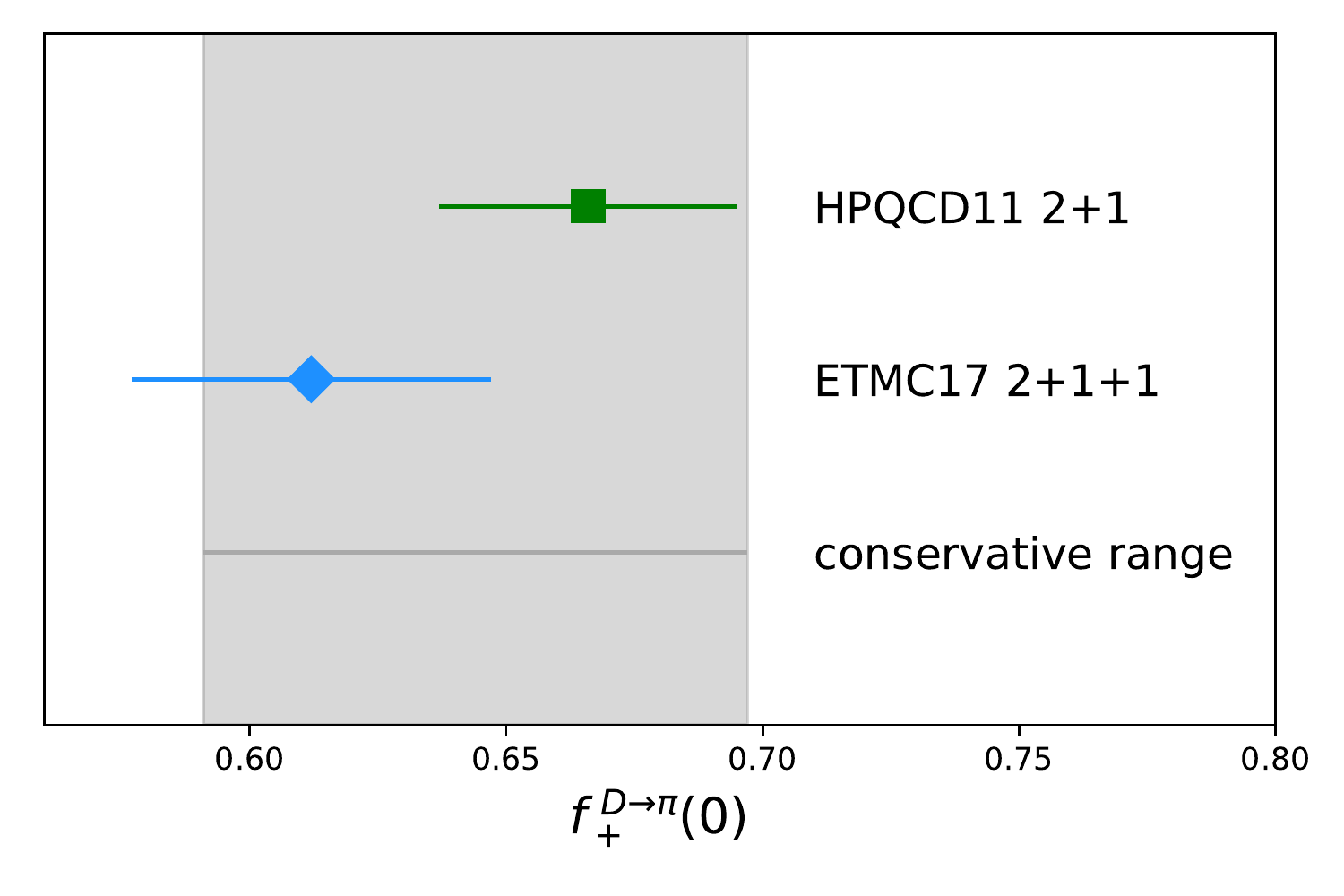}}
\caption{\label{fig:fplus_Dpi_compare} Results for $f_+^{D\to\pi}(0)$.  The green square used staggered fermions with $2+1$ flavors of sea quarks \cite{Na:2011mc} while the blue diamond
used twisted-mass fermions with $2+1+1$ flavors of sea quarks \cite{Lubicz:2017syv}. For the purpose of simplifying the summary plot (Fig.~\ref{fig:VcsVcd}) I display a band which covers
both results.}
\end{figure}

The form factor $f_+(0)$ for $D\to \pi$ provides a normalization for the corresponding semileptonic decay.  There are two results of comparable precision.  The one by HPQCD \cite{Na:2011mc} is on previous generation MILC lattices with $n_f=2+1$, while a more recent one by ETM \cite{Lubicz:2017syv} is on their $n_f=2+1+1$ twisted mass configurations.  The central values differ by about 10\%, which is $2\sigma$ (Fig.~\ref{fig:fplus_Dpi_compare}).  As estimated earlier, the 1-2\% error of quenching the charm is not the likely explanation for this 10\% discrepancy.  Therefore, it is safest to take a conservative estimate which covers both results; see Figure~\ref{fig:fplus_Dpi_compare}.
The picture here will improve if the preliminary $n_f = 2+1+1$ result from Fermilab/MILC  \cite{Li:2019phv} of $f_+^{D\to\pi}(0)$ = 0.625(17)(13) is confirmed.  With the present uncertainties, the value of $|V_{cd}|$ inferred from $D\to \pi\ell\nu$ is consistent with that from $D\to \ell \nu$ no matter which lattice result one takes, and the leptonic determination is what dominates any fit.

In the case of $D \to K \ell \nu$, the HPQCD result \cite{Na:2010uf} for the $q^2=0$ form factor $f_+(0)$ obtained on MILC's $n_f = 2+1$ staggered fermion ensemble is in good agreement with the ETMC result on their $n_f=2+1+1$ twisted mass quark ensemble \cite{Lubicz:2017syv}.  
Preliminary results for the $D\to K$ form factor at $q^2 = 0$, using MILC's 2+1+1 flavor lattices, were also reported by Fermilab/MILC \cite{Li:2019phv}.

The shape of the form factors can of course be determined from lattice QCD, complementing the inferences drawn from measurements of differential branching fractions.  Fitting the experimental data jointly with lattice form factors over a range in $q^2$ has been found to lead to a reduction in the uncertainty of $|V_{cs}|$.  For example the HPQCD found using MILC's staggered $2+1$ lattices  $|V_{cs}| = 0.963(5)_{\mathrm{expt}}(14)_{\mathrm{latt}}$ \cite{Koponen:2013tua}, and the ETM collaboration result using $2+1+1$ flavors of twisted mass fermions leads to $|V_{cs}| = 0.970(33)$ \cite{Riggio:2017zwh}. (In fact ETM also have a result for $|V_{cd}| = 0.2341(74)$ obtained similarly \cite{Riggio:2017zwh}.)

A new result by HPQCD  has recently appeared \cite{Chakraborty:2021qav}, calculating the $D\to K$ form factor over the whole kinematic range with the HISQ valence action on the MILC $2+1+1$ flavor lattices.  They quote an uncertainty on $|V_{cs}|$ below 1\% (when combining their errors quadratically); at this level of precision electromagnetic corrections are important to include.

$|V_{cs}|$ can also be inferred from baryon decays, for example Meinel's $\Lambda_c \to \Lambda \ell^+ \mu_\ell$ form factors \cite{Meinel:2016dqj} combined with BESIII branching fractions \cite{Ablikim:2015prg,Ablikim:2016vqd}. 

HPQCD recently completed form factor calculations for $B_c \to B_{d,s} \ell \nu$ decay \cite{Cooper:2020wnj}.  When experimental measurements are made of these branching fractions, this will lead to a novel method for determining $|V_{cd}|$, $|V_{cs}|$, or their ratio -- one where the $b$ quark is a spectator.  In addition to the novelty of the spectator $b$, this paper is the first to jointly analyze correlation functions obtained with both NRQCD and heavy-HISQ formulations for the $b$ quark.

\subsubsection{Summary}

\begin{figure}
\centering
\resizebox{0.95\hsize}{!}{\includegraphics{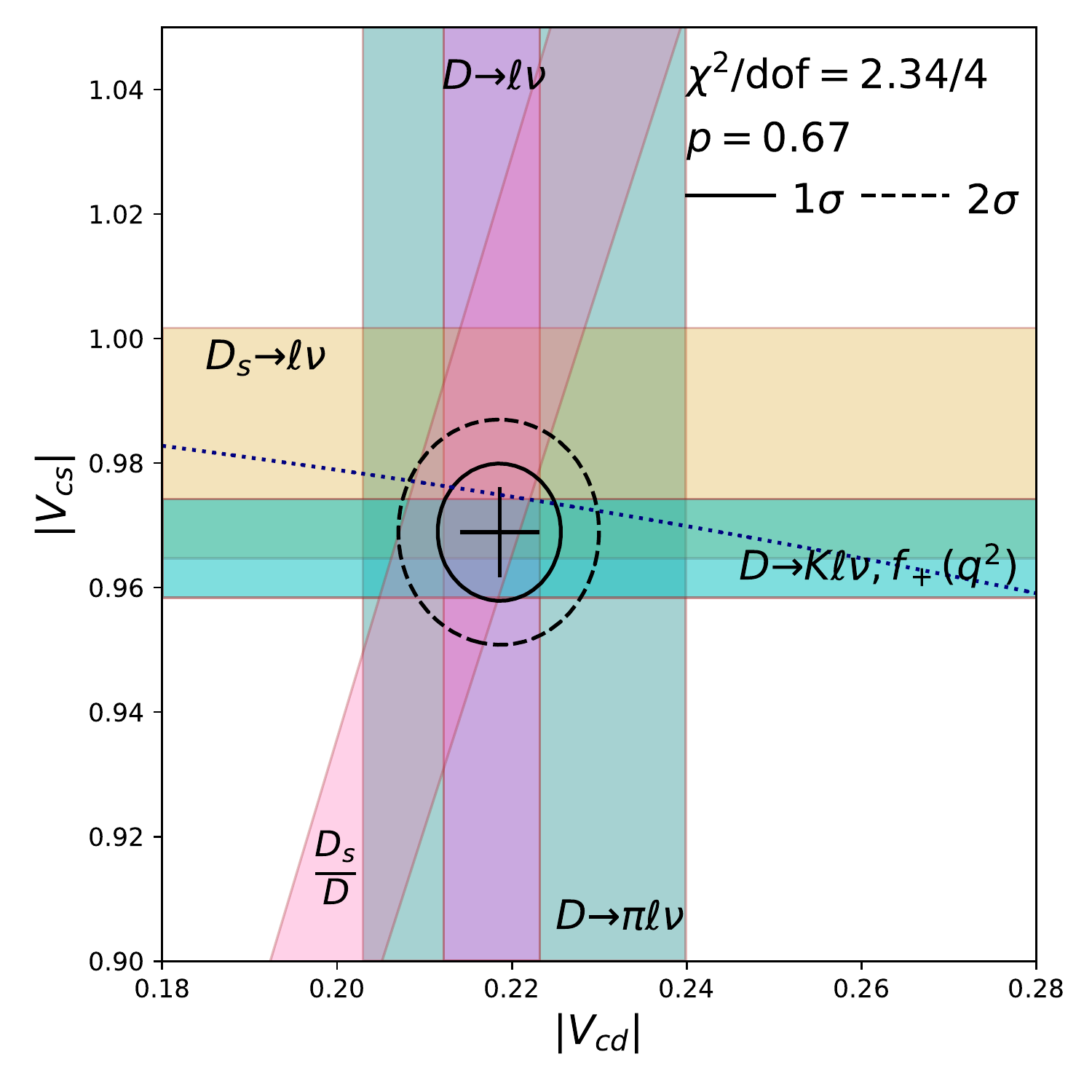}}
\caption{\label{fig:VcsVcd} Constraints in the $|V_{cd}|$-$|V_{cs}|$ plane.  The experimental averages are taken from PDG \cite{Zyla:2020zbs} for the leptonic decays and HFLAV \cite{Amhis:2019ckw} for the semileptonic decays.  A good fit is obtained to constraints from $D\to \ell\nu$ and $D_s\to\ell\nu$ \cite{Bazavov:2017lyh} (including their correlations), and the ratio of decay constants \cite{Boyle:2018knm} as well as $D\to \pi\ell\nu$ \cite{Na:2011mc,Lubicz:2017syv} and $D\to K\ell\nu$ \cite{Chakraborty:2021qav}. The dotted line depicts the values of $|V_{cd}|$ and $|V_{cs}|$ consistent with second-row unitarity.}
\end{figure}


Figure~\ref{fig:VcsVcd} shows the constraints discussed above on $|V_{cd}|$ and $|V_{cs}|$.  A fit to these yields
\begin{align}
|V_{cd}| & = 0.219(5)\\
|V_{cs}| & = 0.969(7) 
\end{align}
and is perfectly consistent with second-row unitarity.
The second row heroes seem to have made several complete and successful journeys.  The discoveries they have brought home have not been revolutionary, but they are nonetheless impressive in their precision and consistency.  Their stories offer hope that the struggles of other heroes still at sea can be overcome.

\subsection{Third row}

The third row heroes have often been the oddsmakers' favorite to return from the abyss triumphant with discovery of New Physics.  In large part this is because direct determinations of $|V_{td}|$ and $|V_{ts}|$ come from loop-mediated processes in the Standard Model.  This SM suppression leaves room for BSM physics to reveal itself.  The most precise measurements here are of the neutral $B^0$ and $B_s^0$ meson mass differences, respectively $\Delta M_d$ and $\Delta M_s$, measured as oscillation frequencies.  In the following I use the experimental averages from PDG 2018 \cite{Tanabashi:2018oca}.

In the past few years, new lattice results have been forthcoming, not only for the matrix element needed for Standard Model predictions of the mass differences, but for matrix elements of the full set of five dimension-6 operators which enter the $\Delta B=2$ effective Hamiltonian.  These come from lattice ensembles with sea quark content $n_f = 2$ (ETM \cite{Carrasco:2013zta}), $2+1$ (Fermilab/MILC \cite{Bazavov:2017lyh}), and $2+1+1$ (HPQCD \cite{Dowdall:2019bea}).   The ETM calculation uses the twisted-mass formulation for all quarks.  The other two groups use staggered fermions (Asqtad and HISQ, respectively) for the light and strange, and the Fermilab or NRQCD formulations, respectively for the bottom quark. There is also a recent calculation of the SU(3)-breaking ratios by the RBC/UKQCD collaboration \cite{Boyle:2018knm}, using domain wall quarks (for all flavors) on configurations with $n_f=2+1$ flavor of sea quarks.

Within uncertainties, there is generally good agreement for the matrix elements, except that the ETM results for the two $(S-P)(S+P)$ operators (color-diagonal and color-mixed) are low compared to the other results.  While the size of the effect of quenching the strange quark is not known, the discrepancy could also be due to the specific RI-MOM scheme used to compute renormalization factors.   A similar discrepancy in $K^0-\bar{K}^0$ operators was recently studied and resolved \cite{Garron:2016mva,Boyle:2017skn}.  Very briefly, condensate contributions can contaminate the gauge-fixed Green's functions, but the infrared behavior can be better controlled with careful choice of kinematics.

\begin{figure}
\centering
\resizebox{0.95\hsize}{!}{\includegraphics{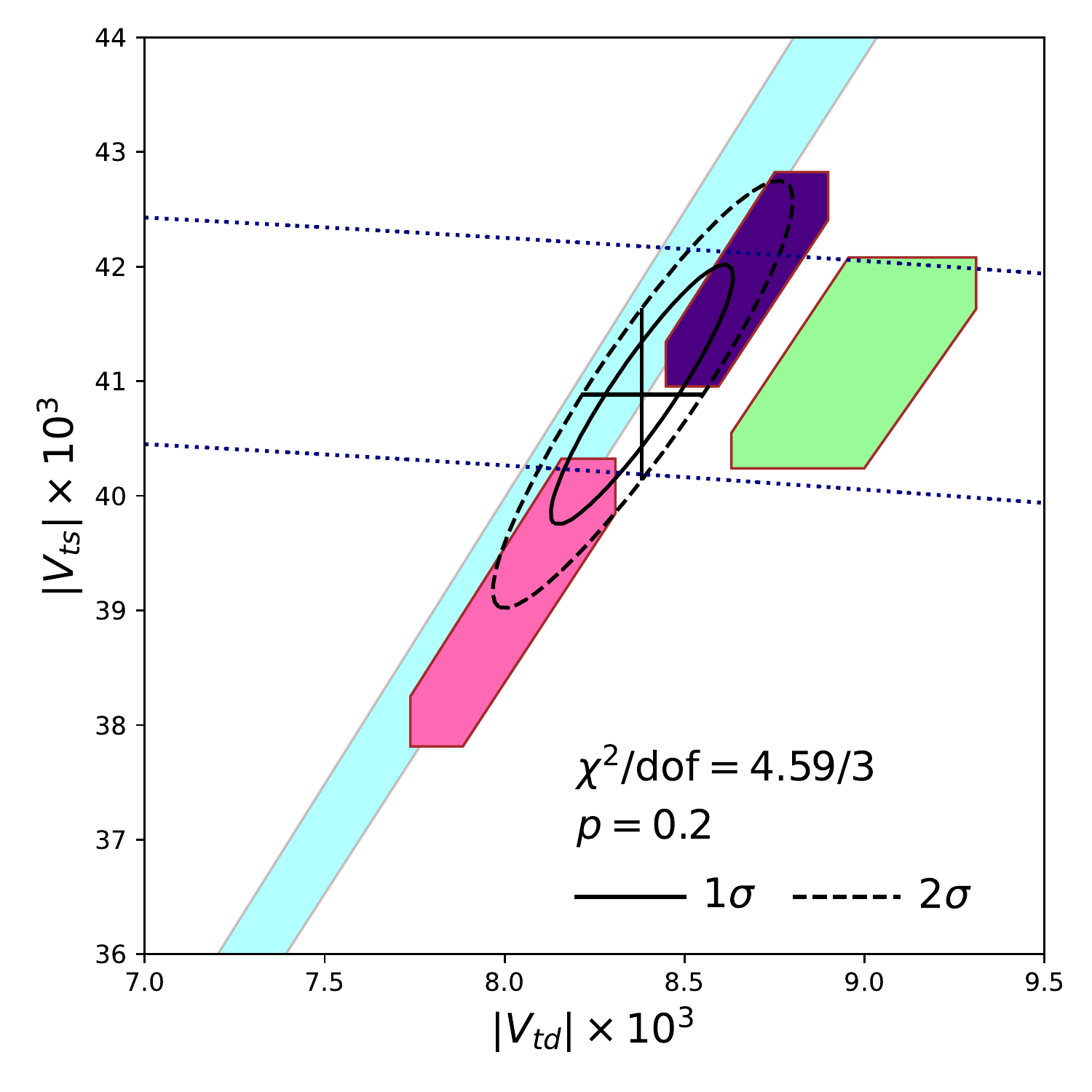}}
\caption{\label{fig:VtdVts} Constraints (at $1\sigma$) from $B^0_{(s)} - \bar{B}^0_{(s)}$ mixing. Cyan diagonal band from RBC/UKQCD \cite{Boyle:2018knm}, pink hexagon from Fermilab/MILC \cite{Bazavov:2017lyh}, indigo hexagon from HPQCD \cite{Dowdall:2019bea}.  The black contours indicate a fit to these 3 results.  For comparison, the green hexagon shows the region constrained by a global CKM fit to tree-level observables \cite{CKMfitter:2018unp} (see \cite{Dowdall:2019bea}). The dotted blue lines indicate the $1\sigma$ region consistent with unitarity using only $|V_{tb}|$.}
\end{figure}

Figure~\ref{fig:VtdVts} shows what the lattice results (with the strange quark unquenched), combined with the experimental measurements, imply for $|V_{td}|$ and $|V_{ts}|$.   Because of correlations and canceling uncertainties, the SM predictions for the ratio $\Delta M_s/\Delta M_d$ is more precise than for numerator or denominator individually.  Hexagons correspond to $1\sigma$ variations in $|V_{td}|$ and $|V_{ts}|$ and their ratio.  I have performed a fit to $|V_{ts}|$ and $|V_{ts}/V_{td}|$ from each group (where available), with $1\sigma$ and $2\sigma$ contours shown, obtaining
\begin{align}
|V_{td}| & = 8.38(17) \times 10^{-3}\\
|V_{ts}| & = 40.9(8) \times 10^{-3} \\
|V_{ts}/V_{td}| & = 4.88(4) \,.
\end{align}

The $\Delta B=2$ matrix elements have recently also been computed using sum rules \cite{King:2019lal}, in very good agreement with lattice results.

In addition to matrix elements of the dimension-6 operators, the SM prediction for the $B_s-\bar{B}_s$ width difference  $\Delta \Gamma_s$ has been improved by having lattice-determined matrix elements of dimension-7 operators \cite{Davies:2019gnp}.

\subsection{Third column}

The $|V_{xb}|$ heroes have been adventuring in the unknown for a long time.  They send messages home, sometimes full of hope and other times of puzzlement.  They have suffered from bouts of infighting which have thankfully subsided, aided by regular peace talks and occasional joint ventures

\subsubsection{Semileptonic $b \to u$ decays}

The most precise determination of $|V_{ub}|$ using an exclusive semileptonic decay comes from $B\to \pi \ell \nu$. 
Because of the expense to extrapolate or compute with the final-state meson at the physical quark mass point, progress takes time.  The complete lattice calculations ready to be included in global averages date from 2015 or earlier. The FLAG review \cite{Aoki:2019cca} includes lattice results on $n_f = 2+1$ lattices of Ref.\ \cite{Gulez:2006dt,Bailey:2015tia,Flynn:2015mha}.  These use either staggered fermions or domain wall fermions for the light quark, and NRQCD, the Fermilab formulation, or the Columbia RHQ action for the $b$ quark.  All of those results come from ensembles of gauge field configurations including the effects of $2+1$ flavors of sea quarks.  Using the BCL parametrization \cite{Bourrely:2008za}, FLAG perform a joint fit to lattice form factor data and binned experimental differential decay rates from BaBar \cite{BABAR:2010uj,Lees:2012vv} and Belle \cite{Ha:2010rf,Sibidanov:2013rkk}.  Their result is
\begin{align}
|V_{ub}| = 3.73(14)\times 10^{-3} \,.
\end{align}

Progress on the next generation calculations is evident.
HPQCD have used the NRQCD-HISQ formulation with MILC HISQ lattices, including physically light pion masses, to show that the soft pion theorem $f_+(q^2_{\mathrm{max}}) = f_B/f_\pi$ holds in the chiral limit \cite{Colquhoun:2015mfa}.
Last year JLQCD presented preliminary results for $B\to \pi$ form factors using M\"obius domain wall fermions for all quarks, with the systematic uncertainties still being quantified \cite{Colquhoun:2019tyq}. 

A new $b \to u$ decay which could be measured by LHCb is $B_c \to D\ell\nu$ \cite{Khanji:2020implications}.
HPQCD is working on $B_c \to D \ell\nu$ form factors, with the rare $B_c \to D_s$ form factors as warm up \cite{Cooper:2020implications}.

Gambino \& Hashimoto have a proposal to address inclusive decays on the lattice \cite{Gambino:2020crt}. Given the longstanding discrepancy between inclusive and exclusive determinations of $|V_{ub}|$, any new line of attack is welcome.

In addition to the vector and scalar form factors, the tensor form factor has also now been computed, allowing complete a SM prediction for the rare decay $B\to \pi \ell^+ \ell^-$ \cite{Bailey:2015nbd}, currently in agreement with the first experimental observation \cite{Aaij:2012de}.

\subsubsection{Leptonic $b \to u$ decay}

The $B$ meson decay constant has been computed by many groups over the years.  FLAG provides a useful summary \cite{Aoki:2019cca}.  Since 2013 the most precise results have been obtained on lattices with 2+1+1 flavors of sea quarks.  In 2017 there was a significant reduction in the uncertainty by the Fermilab/MILC lattice collaborations \cite{Bazavov:2017lyh}.  This reduction comes from using the HISQ formulation for all quark flavors; the renormalization of the lattice axial current is determined fully nonperturbatively.  The Fermilab/MILC results are in good agreement with other 2+1+1 flavor calculations \cite{Dowdall:2013tga,Bussone:2016iua,Hughes:2017spc}.  Of these other determinations, two use nonrelativistic $b$ quarks on MILC's HISQ-action lattices, while the third uses the twisted mass action for all quarks. Thus, we see consistency among the approaches, extrapolating HISQ and twisted quark actions to the physical $b$ limit compared to use of NRQCD.
Any weighted average of the results is dominated by Fermilab/MILC \cite{Bazavov:2017lyh}.

The experimental results for $B^- \to \tau^-\bar{\nu}$ are not very precise presently, in large part due to the small branching fractions and the difficult final state.  BaBar \cite{Aubert:2009wt,Lees:2012ju}  and Belle \cite{Adachi:2012mm,Kronenbitter:2015kls} both have results with two tagging methods, hadronic and semileptonic. However the experiments disagree with each other, with Belle results lower and BaBar results higher.  None of these measurements have reached $5\sigma$ significance.  FLAG has averaged these results, inflating the uncertainty to account for the discrepancy \cite{Aoki:2019cca}, yielding a 30\% determination of the branching fraction.  The $|V_{ub}|$ they infer from this average along with the $2+1+1$ flavor $f_B$ is $|V_{ub}| = 4.05(3)(64)\times 10^{-3}$.  With this large uncertainty, this mode does not yet add much information about $|V_{ub}|$.  Belle II aims to reduce the uncertainty on the $B\to \tau\nu$ branching fraction to approximately 15\% with 5 ab${}^{-1}$ of data and to 5\% with 50 ab${}^{-1}$ \cite{Kou:2018nap}, so there is hope for the future.

\subsubsection{Semileptonic $b \to c$ decays}

From a lattice QCD point of view, the most straightforward route to $|V_{cb}|$ is through $B\to D \ell \nu$ decay.  The initial and final pseudoscalar states are stable to strong interactions, and the Standard Model prediction for the decay, with $\ell = e$ or $\mu$, requires just one form factor, $f_+(q^2)$. Lattice data come from \cite{Bailey:2015rga,Na:2015kha}. FLAG \cite{Aoki:2019cca} combines these results with experimental data in a joint fit to infer $|V_{cb}| = 0.0401(10)$ from $B\to D\ell\nu$ decay.

Experimentally, however, $B\to D\ell\nu$ decay measurements are susceptible to relatively large systematic uncertainties due to the background from $B\to D^* \ell \nu$, with the $D^*$ subsequently decaying to $D\pi$.  Using experimental data for $B\to D^* \ell \nu$ has been the more precise method of determining $|V_{cb}|$ from exclusive decays.  Until recently, the experimental data for the differential decay rate and angular variables has been fit to kinematic functions known as the CLN parametrization \cite{Caprini:1997mu}.  In this case the only information needed from lattice QCD is the normalization, taken from calculations of the axial vector current matrix element at the zero-recoil kinematic point, $h_+(1)$.  Although the $D^*$ decays strongly, its width is narrow and heavy meson chiral perturbation theory gives some guidance.

There was some excitement when Belle published some unfolded data \cite{Abdesselam:2017kjf}, allowing the community to try different kinematic parametrizations. For a time it looked like removing some assumptions present in CLN, by using the BGL parametrization, for example, would resolve the tension between the $B\to D^*\ell \nu$ and inclusive decay determinations of $|V_{cb}|$  \cite{Bernlochner:2017jka,Bigi:2017njr,Grinstein:2017nlq,Bigi:2017jbd,Jaiswal:2017rve,Bernlochner:2017xyx}.  However, the tension remains today.
For a more detailed, recent review of the $|V_{cb}|$ puzzle see \cite{Gambino:2019sif}.

 Progress will be made with new lattice and experiment produce new results. 
 Lattice collaborations are aiming for a full set of form factors for $B\to D^*\ell\nu$ decay, so that the $q^2$ dependence can informed by both lattice and experimental data \cite{Aviles-Casco:2019zop,Kaneko:2019vkx}.  
Because the light spectator quark is expensive, results for decays with heavier spectators, i.e.\  $B_c \to (J/\psi) \ell \nu$ \cite{Harrison:2020gvo} and $B_s \to D^*_s \ell \nu$ \cite{Harrison:2017fmw,Aaij:2020xjy} form factors are milestones along the way.

In addition to being able to infer CKM matrix elements, one also wants to firm up the Standard Model predictions \cite{Bailey:2015rga,Na:2015kha,Harrison:2020nrv} for lepton flavor universality violating ratios \cite{Lees:2012xj,Huschle:2015rga,Aaij:2015yra,Abdesselam:2016cgx,Aaij:2017uff,Hirose:2017dxl,Aaij:2017tyk}.  That said, resolution or confirmation of those anomalies is more likely to occur from reducing experimental uncertainties.

There is no prospect for measuring leptonic $B_c$ decays in the near future.  Nevertheless, lattice calculations of the decay constant $f_{B_c}$ are still welcome.   HPQCD, using heavy HISQ $b$ and $c$ on MILC's 2+1 flavor lattices, found $f_{B_c} = 427(6)$ MeV \cite{McNeile:2012qf} and, while using NRQCD $b$ and HISQ $c$ on MILC's 2+1+1 flavor HISQ lattices, found 434(15) MeV \cite{Colquhoun:2015oha}. In 2018 the European Twisted Mass Collaboration gave a preliminary value for $f_{B_c}$ \cite{Becirevic:2018qlo} which is $2\sigma$ lower than the HPQCD results; however, a proper comparison awaits their finalized result.
It would be good for $f_{B_c}$ to be computed using other actions on other configurations. These allow a test of heavy quark formulations among other things, and is one of the simplest matrix elements involving a $\bar{c}\Gamma b$ current which can be computed in lattice QCD.

\subsubsection{Ratios}

A few years ago saw a novel determination of the ratio $|V_{ub}/V_{cb}|$ using the ratio of $\Lambda_b$ decays  $\Lambda_b \to p\ell\nu$ relative to $\Lambda_b\to \Lambda_c\ell\nu$ \cite{Aaij:2015bfa,Detmold:2015aaa}.  While the experimental and lattice errors are comparable, the lattice determinations could be improved by calculations with larger volumes and lighter sea quark masses.

\begin{figure}
\centering
\resizebox{0.95\hsize}{!}{\includegraphics{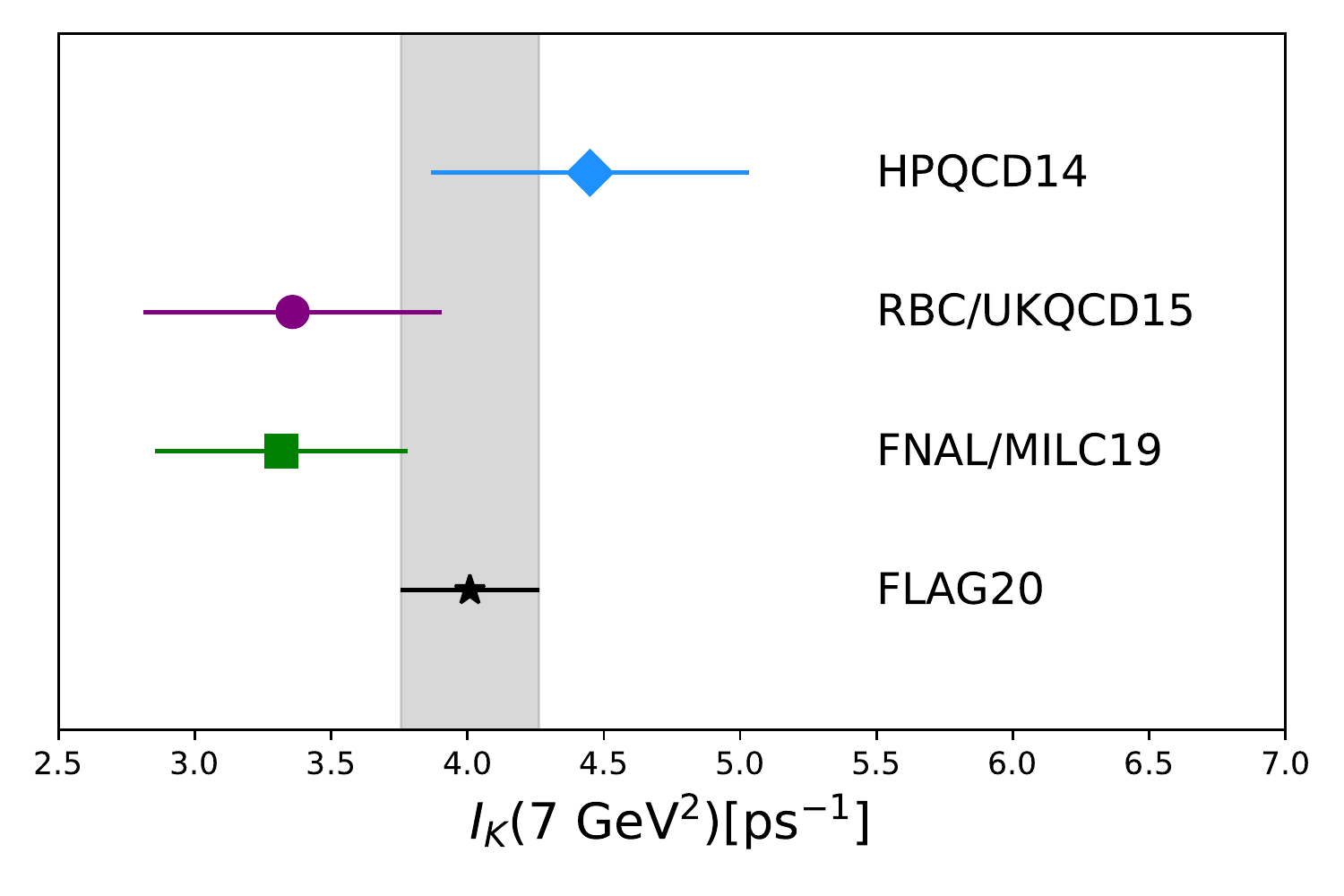}}
\caption{\label{fig:Bs2K_integrated} $B_s\to K \mu\nu$ decay rate for $q^2 \geqslant 7~\mathrm{GeV}^2$ as determined from lattice form factors  \cite{Bouchard:2014ypa,Flynn:2015mha,Bazavov:2019aom} and the FLAG combined fit (black star and grey band) \cite{FLAG:2020update}.}
\end{figure}

Very recently LHCb measured $B_s^0 \to K^-\mu^+\nu_\mu$ relative to $B_s^0 \to D_s^-\mu^+ \nu_\mu$.\cite{Aaij:2020nvo}. Combined with corresponding form factors from lattice QCD (or sum rules) leads to another determination of $|V_{ub}/V_{cb}|$. 
The $B_s \to K\ell\nu$ form factors have been calculated by several groups: FNAL/MILC \cite{Bazavov:2019aom}, HPQCD \cite{Bouchard:2014ypa}, and RBC/UKQCD \cite{Flynn:2015mha}, and $B_s \to D_s\ell\nu$ most recently by HPQCD  \cite{McLean:2019qcx}.  (RBC/UKQCD presented a preliminary update recently \cite{Flynn:2020nmk}.) For the $B_s\to K\mu \nu$ decay, LHCb have divided the branching fraction into two bins.  Since the lattice data are most reliable at low recoil, I focus on the $q^2 > 7~\mathrm{GeV}^2$ bin.  The quantity needed from lattice is thus
$I_K(7~\mathrm{GeV}^2)/I_{D_s}(m_\mu^2)$ where
\begin{align}
I_{P}(t_{\mathrm{cut}}) = \frac{1}{|V_{xb}|^2}\int_{t_{\mathrm{cut}}}^{t_-} dt\, \left.\frac{d \Gamma(B_s \to P \ell \nu)}{dt} \right|_{\mathrm{SM}}
\end{align}
and $t_- = (M_{B_s} - M_P)^2$.  
Fig.~\ref{fig:Bs2K_integrated} shows the results using the form factor fits from each of the three collaborations.  FLAG \cite{FLAG:2020update} has combined the lattice data and found they can obtain a good fit, despite the tension apparent in Fig.~\ref{fig:Bs2K_integrated}.  

One should keep in mind that the spatial momentum of the $K$ in the lattice correlation functions generally corresponds to $q^2 \geqslant 17~\mathrm{GeV}^2$.  Thus the integration down to 7 GeV${}^2$ involves an extrapolation of the lattice fits.  In the case of calculations using nonrelativistic $b$ quarks \cite{Bouchard:2014ypa}, this extrapolation goes in the direction where the effective theory is worsening.  The underlying assumption of a clear separation of scales between QCD physics and the $b$ quark mass.  This has been shown to work well for many matrix elements, including form factors in the low recoil regime, where $\Lambda_{\subrm{QCD}}/m_b \approx 0.1$ is the relevant ratio.  However, as $q^2$ is decreased from $17 ~\mathrm{GeV}^2$ to $7 ~\mathrm{GeV}^2$  the spatial kaon momentum increases from 1 GeV to 2 GeV.  This means that $|\mathbf{p}_K|/m_b$ is growing, spoiling the separation of scales.  Operators which may be negligible in the low recoil regime can develop large matrix elements.  For example $J_4 = \frac{1}{2m_b} \bar{q} \bfnabl_k Q$ is neglected in NRQCD calculations since it only enters at $O(\alpha_s)$. However, the corresponding matrix elements grow as the kaon momentum increases and cannot be neglected far away from the low recoil regime \cite{Gulez:2006dt}.  

The fact that FLAG find a good combined fit to the published lattice data suggests that the quoted lattice uncertainties hold well in the kinematic regime where the calculations are done.  Further investigation into the extrapolations to lower $q^2$ is needed.  In addition to issues with using NRQCD away from low recoil, there are different methods for disentangling the physical $q^2$ dependence of the form factors from lattice spacing and quark mass dependencies.  These should be yield consistent results, but if they do not, then this needs to be understood.  Finally, greater scrutiny could be given to the effect of imposing the constraint $f_+(0) = f_0(0)$ in fits to the lattice data.  Since we expect errors to grow as we extrapolate from high to low $q^2$, the fits should tolerate some deviation from this equality.  After all, in most cases we are more interested in the accuracy of the form factors in the medium-to-large $q^2$ region than in tightly enforcing equality at $q^2 = 0$.  

Since lattice data are being used for both  $B_s \to K\ell\nu$ and $B_s \to D_s \ell\nu$ form factors, it is sensible to ask whether there are correlations which need to be taken into account.
This ratio of decays $B_s \to K\ell\nu$ to $B_s \to D_s \ell\nu$ was specifically addressed in \cite{Monahan:2018lzv}.  They performed a simultaneous fit to their lattice data, publishing the full covariance matrix for their form factor shape parameters.  Using these to compute $I_K(7~\mathrm{GeV}^2)/I_{D_s}(m_\mu^2)$ gives a 22\% correlation between numerator and denominator.

\subsubsection{Summary}

\begin{figure}
\centering
\resizebox{0.95\hsize}{!}{\includegraphics{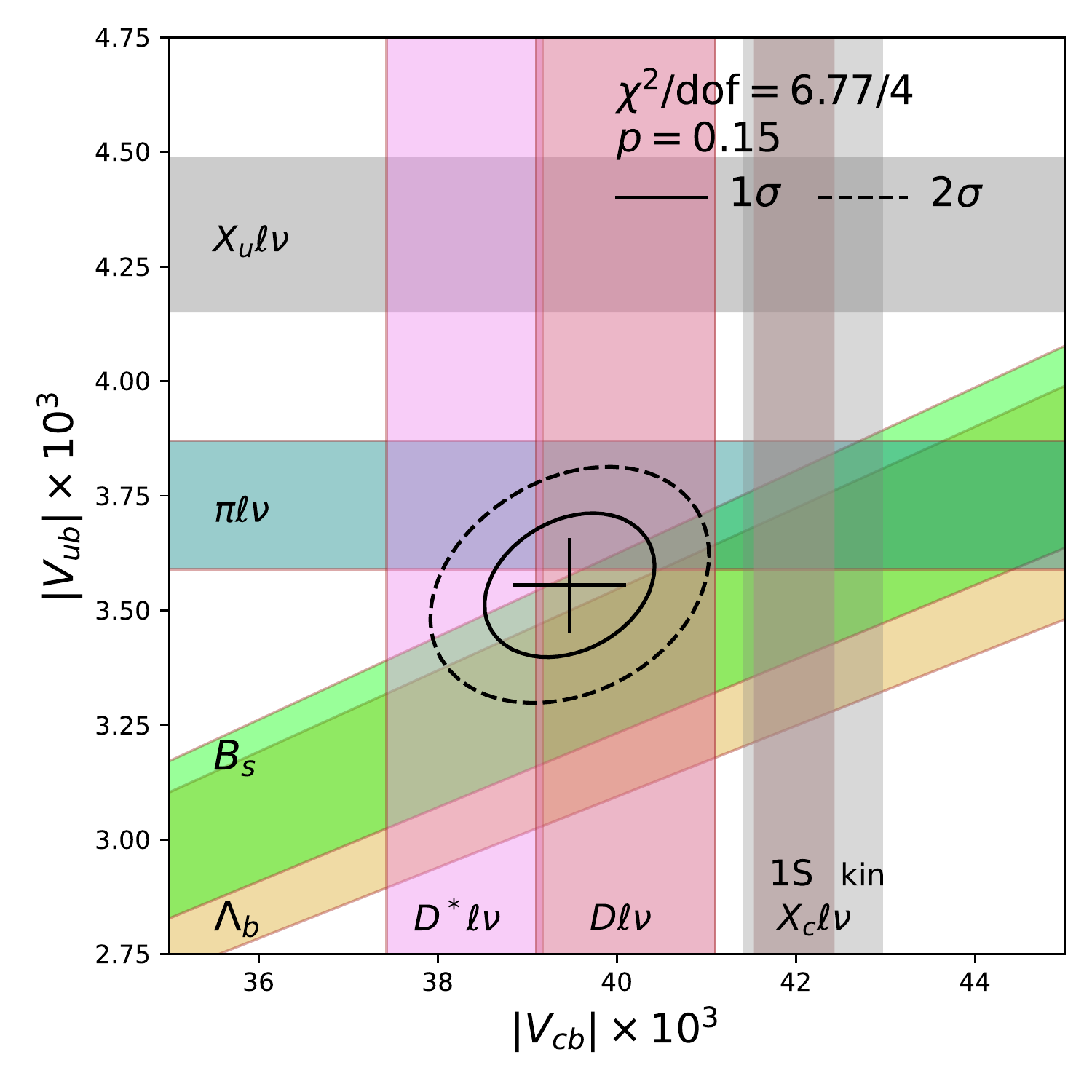}}
\caption{\label{fig:VubVcb} Constraints in the $|V_{cb}|$-$|V_{ub}|$ plane as discussed in the text.  The fit indicated is to the data and SM theory for exclusive modes, including $B\to \tau \nu$ (not shown). Gray bands show the inclusive determinations taken from HFLAV 2019 \cite{Amhis:2019ckw}; the $|V_{ub}|$ result in the recommended GGOU scheme and both 1S and kinematic schemes for $|V_{cb}|$.}
\end{figure}

Figure \ref{fig:VubVcb} summarizes the situation with the third column of the CKM matrix.
the solid (dashed) curve depicts the $1\sigma$ ($2\sigma$) contours.  A decent fit is obtained when including the constraints from the exclusive semileptonic decays (as well as the weak constraint from leptonic $B$ decay). The fit results are
\begin{align}
|V_{cb}| & =  39.4(6) \times 10^{-3} \\
|V_{ub}| & = 3.61(12) \times 10^{-3} \,.
\label{eq:VubVcbfit}
\end{align}
Unfortunately we cannot expect our $|V_{xb}|$ heroes to return until the discrepancy with the determinations from inclusive decays is better understood.

\section{Rare processes}
\label{sec:rare}

\subsection{Rare kaon processes}

\subsubsection{Direct CP violation in $K\to \pi\pi$}

Consider the decays of neutral kaons to two pions.  The weak, or flavor, eigenstates $|K^0\rangle$ and $|\bar{K}^0\rangle$ map into each other under a CP transformation.  One can form CP eigenstates $| K_1 \rangle$ and $|K_2\rangle$ as linear combinations of the weak eigenstates.  Only the CP-even eigenstate can decay to two pions.

The mass eigenstates $|K_S\rangle$ and $|K_L\rangle$ are not pure CP eigenstates.  One is longer-lived than the other, with the $K_S$ decaying almost always to $\pi^+ \pi^-$ or $\pi^0\pi^0$ and the $K_L$ decaying to a variety of other modes, notably to 3 pions.  However, the $K_L$ also occasionally decays to 2 pions.
The relevant ratios of matrix elements of the weak Hamiltonian $H_W$ are defined as follows:
\begin{align}
\eta_{+-} & = \frac{\langle \pi^+ \pi^-|H_W | K_L\rangle}{\langle \pi^+ \pi^-|H_W | K_S\rangle} \approx \varepsilon + \varepsilon' \nonumber \\
\eta_{00} & = \frac{\langle \pi^0 \pi^0|H_W | K_L\rangle}{\langle \pi^0 \pi^0|H_W | K_S\rangle} \approx \varepsilon - 2\varepsilon'  \,.
\end{align}
If the only source of CP violation came from mixing, then we would have $\eta_{+-} = \eta_{00} \approx \varepsilon$, so $\varepsilon'$ quantifies what is called ``direct CP violation.'' In fact experiments measure \cite{Batley:2002gn,Abouzaid:2010ny,Zyla:2020zbs}
\begin{align}
\mathrm{Re} \frac{\varepsilon'}{\varepsilon} \approx  \frac16 \left( 1 - \left|\frac{\eta_{00}}{\eta_{+-}}\right|^2\right) = 1.66(23) \times 10^{-3} \,.
\label{eq:epsilonprime_expt}
\end{align}
(The approximation symbols above indicate truncation of higher-order terms in $(\varepsilon'/\varepsilon)$, which are presently negligible.)

In terms of the amplitudes for decay into specific isospin states $I = 0$ or 2, $A_I \equiv \langle(\pi \pi)_I | H_W | K^0\rangle$, the direct CP violating parameter can be determined through
\begin{align}
\varepsilon' &= \frac{i}{\sqrt2}\frac{\mathrm{Re}A_2}{\mathrm{Re}A_0}
\left(\frac{\mathrm{Im}A_2}{\mathrm{Re}A_2} - \frac{\mathrm{Im}A_0}{\mathrm{Re}A_0}\right)e^{i(\delta_2-\delta_0)}
\end{align}
where the difference in the scattering phase shifts $\delta_2 - \delta_0 \approx -\pi/4$.

RBC/UKQCD have recently published a new result for $\varepsilon'/\varepsilon$ \cite{Abbott:2020hxn}. This updates their previous result \cite{Bai:2015nea} which found a value $2\sigma$ below the experimental value.  That work also gave a phase shift for the $I=0$ channel which was in significant disagreement with a dispersive analysis \cite{Colangelo:2001df}.  The updated work substantially improves the analysis of the lattice $\pi\pi$ correlation functions.  By tripling the statistics and using a larger set of interpolating operators, the authors were able to conclude that their previous correlation functions possessed larger-than-expected contamination from excited states.  A new preprint details their updated analysis of the $\pi\pi$ phase shifts \cite{Blum:2021fcp}, which are now in good agreement with the dispersive analysis results.

Their new result is $\mathrm{Re}(\varepsilon'/\varepsilon) = 2.17(26)(62)(50) \times 10^{-3}$, where the uncertainties are respectively due to statistics, isospin-conserving systematic errors, and omitted isospin breaking effects.  This is now compatible with (\ref{eq:epsilonprime_expt}).
They also give a new result for 
\begin{align}
\frac{\mathrm{Re}A_0}{\mathrm{Re}A_2} = 19.9(2.3)(4.4)
\end{align}
where the denominator comes from \cite{Blum:2015ywa}.

These heroes deserve a big banquet and celebration now that they have returned home after an arduous journey.  They should not overindulge, however. The $\varepsilon'/\varepsilon$ saga is not over. The experimental uncertainty is less than 20\%, so we need these heroes to lead a new band on the next adventure.  New lines of attack include a second, finer lattice spacing and improving the operator matching by working with a four-flavor effective theory (instead of one where charm is integrated out) \cite{Tomii:2018zix,Tomii:2019esd}.

\subsubsection{Rare semileptonic $K$ decays}

Members of the RBC/UKQCD collaboration have set off on another ambitious journey, to calculate the long-distance contributions to rare semileptonic decays.
This is part of a program initiated for both $K\to \pi \ell^+\ell^-$ \cite{Christ:2015aha} and $K \to \pi \nu \bar\nu$  \cite{Christ:2016eae} decays, with the basic  approach proposed some time ago \cite{Christ:2010gi}.  In the former case, one needs matrix elements of bilocal products of the effective $s\to d$ Hamiltonian with the electromagnetic current.  In the latter case there are several bilocal operators which are obtained in the appropriate effective field theory.

$K \to \pi \nu \bar\nu$ is predominantly governed by short-distance physics.  However, long-distance effects could contribute to $K^+\to \pi^+ \nu \bar \nu$ at the 5-10\% level.  The NA62 experiment has just reported evidence for this decay at the $3.4\sigma$ level \cite{CortinaGil:2021nts}, and are reported to be aiming for a 10\% measurement eventually \cite{Christ:2019dxu}. It is timely for lattice QCD to determine these nonlocal contributions.  In their recent paper \cite{Christ:2019dxu} have carried out a calculation with quark masses corresponding to nearly physical pion mass.  They investigated several effects, such as momentum dependence, the contributions from disconnected diagrams, and their ability to control unphysical effects.  The results look promising so far, and it looks like they may be able to continue onward toward a determination of the required amplitude, at least at one kinematic point.

\subsection{Rare $\bm{b}$ decays}

With the many new measurements of $b\to s$ decays in the past 5-10 years, there are some exciting deviations from Standard Model predictions.  Among these are a $\approx 3\sigma$ disagreement between experiment and SM theory in the $B\to K^* \mu^+ \mu^-$ angular observable $P_5'$ \cite{Aaij:2015oid,Khachatryan:2015isa} and the ratios $R(K^{(*)})$ \cite{Aaij:2014ora,Aaij:2017vbb,Aaij:2019wad,Aaij:2021vac}, of $B\to K^{(*)}\ell^+ \ell^-$ modes with muon- versus electron-pair in the final state.  Lattice QCD is not needed for these theory calculations. However, LQCD determinations of the $B\to K$ \cite{Bouchard:2013mia,Bouchard:2013eph,Bailey:2015dka}, $B\to K^*$, and $B_s\to \phi$ form factors \cite{Horgan:2013hoa,Horgan:2013pva} help determine the differential decay rates.  The experimental decay rates \cite{Aaij:2014pli,Aaij:2015esa} are a bit lower than the SM predictions taken with the assumption that the lattice form factors are the full story.  It is intriguing that the same extension to the SM effective interaction, namely an enhanced Wilson coupling of the operator $Q_9' \propto (\bar{s}\gamma_\mu P_R b)(\bar{\ell}\gamma^\mu \ell)$ , would both resolve the $P_5'$ discrepancy and lower the theory prediction for these branching fractions.

Unfortunately, the SM predictions are not of the same ``gold'' standard as those discussed in Sec.~\ref{sec:ckm}.  The decay rates above all receive contributions from matrix elements of nonlocal operators, most importantly the product of $Q_2 = (\bar{s}\gamma_\mu P_L c)(\bar{c} \gamma^\mu P_L b)$ with the electromagnetic vector current, which is enhanced when the momentum transfer is equal to the mass of a charmonium resonance (e.g.\ \cite{Lyon:2014hpa}).  There had been some hopes of treating these phenomenologically \cite{Grinstein:2004vb,Khodjamirian:2010vf,Beylich:2011aq,Bobeth:2017vxj}, but not to the accuracy required to discern new physics, given the small size of the discrepancy.  Some preliminary steps using lattice QCD to ask questions about these nonlocal matrix elements have been taken in \cite{Nakayama:2020hhu}.

Furthermore, the $K^*$ and the $\phi$ decay strongly, so the narrow width approximation assumed in \cite{Horgan:2013hoa,Horgan:2013pva} is an uncontrolled approximation, perhaps more valid for the narrow $\phi$ than the broad $K^*$ resonance.  The way forward has been mapped in \cite{Briceno:2014uqa}, which sets out a method for studying the full transition amplitude for $B\to K\pi$.

The rare decay $B \to K^{(*)} \nu \bar\nu$ is short-distance dominated.  This is because of a harder GIM suppression compared to the charged lepton final state.  The contribution from charm quark loops is smaller than top quark loops by a factor $O(10^{-3})$ \cite{Buchalla:1995vs}.  Although the vector meson final state is as problematic here as in the charged lepton mode, the $B\to K$ lattice form factor is enough for a reliable Standard Model prediction of $B^+ \to K^+ \nu \bar\nu$, which could be measured by Belle II \cite{Kou:2018nap}.

\section{Conclusion}
\label{sec:conclusion}

These are just a few tales of the flavor physics heroes.   Each band of adventurers is facing their own set of obstacles based on their chosen route.  For some, discretization errors are more treacherous, for others renormalization factors are a roadblock, and others require vast resources to travel even farther.  With more powerful machines, the need for risky extrapolations is being reduced, although the form factor heroes have the ever-present challenge of ensuring the kinematic shape is
safely interpolated.  One of the themes emerging in 
many of the stories is that isospin-breaking effects are the next challenge which need to be faced.

Results directly testing the CKM paradigm invariably receive the most attention, but there is much work going on behind the scenes as well, improving methods and testing assumptions.
One of the best things about the International Symposia on Lattice Field Theory is the format, where plenary sessions are complemented by many parallel sessions.  Through these and the poster session, one appreciates the breadth and depth of research which advances flavor physics, hadron physics more broadly, and physics beyond the Standard Model.  It was a shame to have lost that in 2020.  Colleagues should be applauded for their efforts to fill the gap with virtual meetings on a smaller scale and with this volume.  All this experience will undoubtedly be applied to making the MIT virtual Lattice 2021 a success.  Even so, I look forward to seeing you in person at the next opportunity, perhaps in Bonn for Lattice 2022 if not sooner. \textit{Bis wir uns wieder treffen.}

\section*{Acknowledgments}

I am grateful to the International Advisory Committee and Local Organizing Committee of Lattice 2020 for the invitation to prepare this review, and for their patience awaiting its completion.  I could not resist including some recent exciting measurements and calculations, so the review reflects the situation circa early 2021. I have had several helpful discussions with Christine Davies and Bipasha Chakraborty during the early stages of preparation.

%
\bibliographystyle{epj}
\bibliography{mbw}

\end{document}